\documentclass[graybox]{svmult}
\usepackage{mathptmx}       
\usepackage{helvet}         
\usepackage{courier}        
\usepackage{type1cm}        
\usepackage{makeidx}         
\usepackage{url}            
\usepackage{graphicx}        
\usepackage{multicol}        
\usepackage[bottom]{footmisc}
\makeindex             

\begin{document}
\title*{Electronic Structure Calculations with LDA+DMFT}
\author{Eva Pavarini}
\institute{Eva Pavarini \at Institute for Advanced Simulation and  Peter Gr\"unberg 
Institut, Forschungszentrum J\"ulich, Germany \email{e.pavarini@fz-juelich.de}}
\maketitle

\abstract{The LDA+DMFT method is a very powerful tool for gaining insight into the physics of
strongly correlated materials.
It combines traditional {\em ab-initio}  density-functional techniques with the 
dynamical mean-field theory. 
The core aspects of the method are (i) building material-specific Hubbard-like many-body models
and (ii) solving them in the dynamical mean-field approximation. 
Step (i) requires the construction of a localized one-electron basis, typically
a set of Wannier functions. It also involves a number of approximations, such as the choice 
of the degrees of freedom for which many-body effects are explicitly taken into account, the
scheme to account for screening effects, or the form of the double-counting correction.
Step (ii) requires the dynamical mean-field solution of multi-orbital generalized Hubbard models.
Here central is the quantum-impurity solver, which is also the 
computationally most demanding part of the full LDA+DMFT approach.
In this chapter I will introduce the core aspects of the LDA+DMFT method and present 
a prototypical application.}

\section{The Strong Correlation Problem}
In the non-relativistic limit electrons in a crystal are typically described  by the Hamiltonian (in atomic units)
\begin{eqnarray}\label{ele}
\nonumber
H&=&-\frac{1}{2} \sum_i \nabla^2_i 
 + \frac{1}{2}\sum_{i \ne {i^\prime}} \frac{1}{|{\bf r}_i-{\bf r}_{i^\prime} |}
- \sum_{i\alpha} \frac{Z_\alpha}{|{\bf r}_i-{\bf R}_\alpha|}  + \frac{1}{2} \sum_{\alpha \ne \alpha^\prime} \frac{Z_\alpha Z_{\alpha^\prime} }{|{\bf R}_\alpha  -{\bf R}_{\alpha^\prime}|}
 \\
  &=&{T}_e+{V}_{ee} +{V}_{en} +{V}_{nn},
\end{eqnarray}
where $\{ {\bf r}_i \}$ are the coordinates of the $N_e$ electrons, $\{ {\bf R}_\alpha \}$
those of the $N_n$  nuclei, $Z_\alpha$ their atomic numbers, and $M_\alpha$ their masses. 
Although it appears innocent, the Schr\"odinger equation $H_e \psi =\varepsilon\psi$ has a simple solution only in the non-interacting electron limit (${V}_{ee}=0$).
In such a case it is sufficient to find the eigenvalues and eigenvectors of the 
one-electron Hamiltonian
\begin{eqnarray*}
{h}({\bf r})=-\frac{1}{2} \nabla^2 - \sum_{\alpha} \frac{Z_\alpha}{|{\bf r}-{\bf R}_\alpha|}
=-\frac{1}{2} \nabla^2+v_{\rm ext}({\bf r}).
\end{eqnarray*}
In a crystal, because of lattice translational invariance,  the eigenvectors of ${h}_e ({\bf r})$ are Bloch functions, 
$\psi_{n{\bf k}\sigma}({\bf r})$, and the eigenvalues band energies, $\varepsilon_{n{\bf k}}$;  
the many-body $N_e$-electron states can be then built from the Bloch states as Slater determinants.
For an interacting system ($V_{ee}\ne0$) we are, however, left in the realm of approximations.

In {\em some} limit the independent-particle picture still holds.
Landau Fermi-liquid theory suggests that, at low enough energy and temperature, the elementary excitations of the interacting Hamiltonian (\ref{ele}) could be described by almost independent Fermionic {quasi particles}, Fermions with heavy masses $m^*$ and finite life-time $\tau^{\rm QP}$ 
\begin{eqnarray*}
\varepsilon_{n {\bf k}}^{\rm QP} &=&\frac{m}{m^*} \varepsilon_{n {\bf k}},\\
 \tau^{\rm QP} &\propto& (aT^{2}+b\omega^{2})^{-1}.
\end{eqnarray*}
Remarkably, a very large number of materials do exhibit low-energy Fermi-liquid behavior,
and a violation of the Fermi-liquid picture is typically an indication that something surprising
is going on.

Starting from a different perspective, using the {\em standard model} of solid state physics, the density-functional theory (DFT) \cite{lda,barth,martin}, one can show  that (\ref{ele}) can be mapped into an auxiliary one-electron problem
(Kohn-Sham equations) where the external potential is replaced by  
\begin{eqnarray*}
v_{\rm R}({\bf r})=- \sum_{\alpha} \frac{Z_\alpha}{|{\bf r}-{\bf R}_\alpha|}+\int d {\bf r}^\prime \frac{n({\bf r}^\prime)}{|{\bf r}-{\bf r}^\prime|}+\frac{\delta E_{xc}[n]}{\delta n}, 
\end{eqnarray*}
and $n({\bf r})$ is the electronic ground-state density.
The first term of $v_{\rm R}({\bf r})$ is the external electron-nuclei interaction, the second is the long-range Hartree interaction and the third is the exchange-correlation potential.
The DFT exchange-correlation functional $E_{xc}[n]$ is universal but also unknown.
Thus, although DFT is in principle an exact ground-state theory, in practice we work with approximated forms of $E_{xc}[n]$.
The most common approximation is the  local-density approximation (LDA), in which $E_{xc}[n]$ is replaced
by its expression for an  interacting homogeneous electron gas, 
\begin{figure}[t]\label{extension}
 \centering
 \rotatebox{0}{\includegraphics[width=0.725\textwidth]{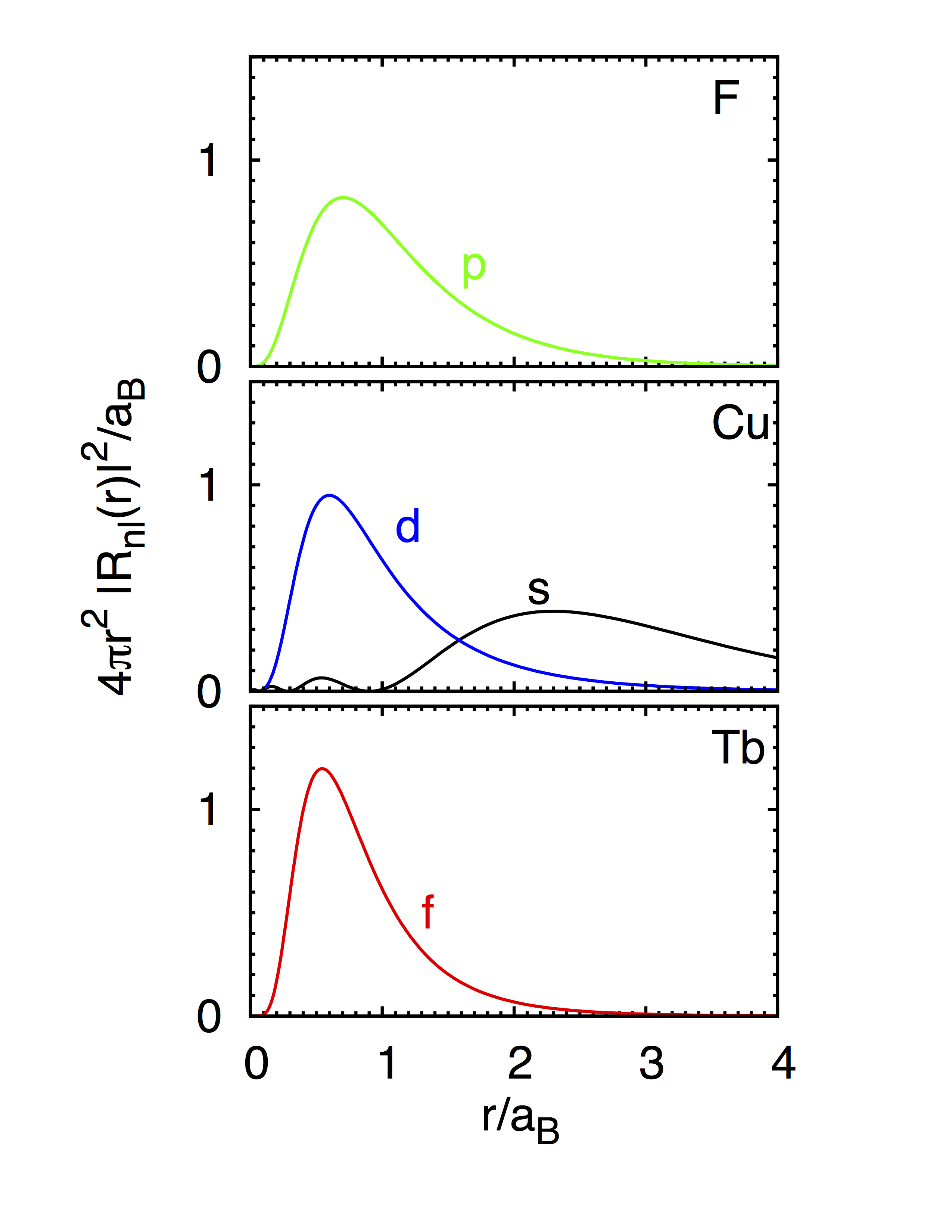}}
 \caption{ Localization of atomic orbitals (calculated in the LDA approximation) in exemplary cases: F ($p$), Cu ($d$ and $s$), and Tb ($f$) atoms. 
 The figure shows $4\pi r^2|R_{nl}(r)|^2/a_B$  as a function of the distance 
 from the position of the nucleus, $r/a_B$. }
 \end{figure}
\begin{equation} \label{lda}
E_{xc}[n] \sim \int d{\bf r} \; \epsilon_{xc}^{\rm LDA}(n({\bf r})) n({\bf r}).
\end{equation}
The LDA appears particularly justified if $n({\bf r})$ varies slowly in space.
In this limit we can think of splitting space into regions in which the density is basically constant
and the system can indeed be described by an interacting homogeneous electron gas with a given 
ground-state electron density $n({\bf r})$. By adding up the contributions 
of all these regions we can obtain expression (\ref{lda}).  

The LDA and its generalizations have opened the path to {\em ab-initio} electronic-structure 
calculations, leading to the astonishing successes of DFT
in explaining and predicting the electronic properties of complex materials \cite{lda}. 
Even if Kohn-Sham eigenergies and orbitals are mathematically only the solution of an auxiliary problem, they proved very useful to describe the electronic structure of materials, even at finite temperature and for excited states. 

This is not, however, the full story. 
For some systems simple approximations to the exchange-correlation functional such 
as the LDA qualitatively fail.\footnote{For further discussion about the exchange-correlation term, see chapters of Tzanov and Tuckerman, Ghiringhelli, Delle Site, Karasiev {\it et al.}, Watermann et {\it al.}; for the specific case of strongly-correlated electrons see the chapter of Malet et {\it al.}.}
These are the so-called {\em strongly correlated systems}, materials in which many-body effects
manifest themselves in the form of emergent co-operative phenomena.  
A paradigmatic example is that of Mott insulators. Because of Coulomb repulsion, several transition-metal compounds with partially filled $d$ shells are experimentally paramagnetic insulators in a large temperature range, despite being described as good metals in LDA. Simple improvements of the LDA functional do not solve the discrepancy. 
Other examples are high-temperature superconducting cuprates, heavy Fermions, Kondo systems, and correlated organic crystals.
The problem is usually the description of many-body effects between {\em localized} electrons from open $d$ or $f$ shells
(see Figure \ref{extension}). Because of the strong Coulomb repulsion, the dynamics of a single electron
depends on the position of all other electrons, and cannot be understood within an independent electron picture, as the one arising from simple approximations to the DFT exchange-correlation functional. 

It is interesting to observe that for several strongly correlated systems the Fermi-liquid picture still holds in the low-energy regime, although the effective masses can reach extreme values, as in heavy Fermions.
One could therefore think of finding some effective potential that yields such very high masses.
The actual energy region of validity of the Fermi-liquid theory for a given system is however 
unknown and it might be very narrow. This is typically the case
for heavy-Fermions. Furthermore Fermi-liquid theory relies on perturbation theory,
and inherently non-perturbative many-body phenomena are known. In
systems such as Mott insulators even the Fermi-liquid picture breaks down.

In the lack of a better option, the study of strong correlations effects has been confined for 
a long time to simple many-body models.  This {\em minimal} approach lead to striking successes, such as understanding the mechanism of the Kondo effect in diluted magnetic alloys, or to important developments in many-body theory, among which also the dynamical mean-field theory (DMFT) can be counted \cite{dmft,rmp}. Nevertheless,  simple models are hardly sufficient to describe the complications of many-body
effects in real materials. Thus during the years various attempts had been made to combine
{\em ab-initio} techniques and many-body methods \cite{Hewson}. 
The breakthrough came with the development of the LDA+DMFT (local-density approximation + dynamical mean-field theory) method \cite{original}.

The LDA+DMFT approach can be roughly split into two  steps. 
The first consists in building
{optimal} material-specific many-body models, exploiting the power of practical DFT. 
In this step we construct a localized {\em single-electron basis} and calculate
the parameters of the many-body model in such a basis. 
Crucial in model building is the identification of the electrons responsible for correlation effects. 
The second step consists in solving the resulting generalized Hubbard-like model with
DMFT or, if possible, its extensions. Here one is faced with other challenges,  
in particular the  solution of the DMFT {\it quantum-impurity problem}. 
Typically this is the computationally most demanding part of the approach.
In this introductory chapter I discuss the two main steps of the LDA+DMFT method,
and I present a characteristic application.
A more extended introduction can be found in Ref.~\cite{eva2011}; more
details on model building or quantum impurity solvers are given  in Refs.~\cite{book2011,book2012}.
Other reviews on the LDA+DMFT method an its successes are Refs.~\cite{georges,georges2,imada,kotliar}.

\section{Material-Specific Many-Body Models from DFT}
The successes of the LDA suggest that the LDA Kohn-Sham orbitals carry the essential information about the structure and bonding of a given material.
Thus they are best suited as a starting point to construct bases for material-specific  many-body models. 
In recent years it has been shown that indeed a successful scheme consists in building localized
Wannier functions $\psi_{in\sigma}({\bf r})$ from LDA  Bloch functions 
\begin{eqnarray*}
\psi_{in\sigma}({\bf r})= \frac{1}{\sqrt N} \sum_{{\bf k}} e^ {-i {\bf R}_i\cdot {\bf k}} \; \psi_{n{\bf k}\sigma}({\bf r}). 
\end{eqnarray*}
Localized Wannier functions can be obtained in different ways. Successful methods are the {\em ab-initio} downfolding procedure based on the NMTO approach \cite{njp,olebook2012}, the maximally-localized Wannier functions algorithm of Marzari and Vanderbilt \cite{mlwf} and projector techniques \cite{projectors}.
In the LDA Wannier basis the many-body Hamiltonian (\ref{ele}) takes the form
\begin{equation} \label{Hele}
{H}={H}_{\rm LDA} +H_{U} -{H}_{\rm DC}.
\end{equation}
The first term, $H_{\rm LDA}$, corresponds to the sum of the kinetic and potential energy ($T_e+V_{en}$), and can be expressed as
\begin{eqnarray*} 
{H}_{\rm LDA}=-\sum_{\sigma}\sum_{i i^\prime}\sum_{mm^\prime} t^{i,i^\prime}_{m,m^\prime} c^\dagger_{i m \sigma} c^{\phantom{\dagger}}_{i^\prime m^\prime\sigma},
\end{eqnarray*}
where $ c^\dagger_{im\sigma}$ ($c^{\phantom{\dagger}}_{i m\sigma}$) creates (destroys) an electron with spin $\sigma$ in orbital $m$ at site $i$, and the elements of matrix are
\begin{eqnarray*}
t^{i,i^\prime}_{m,m^\prime}= -\int d {\bf r} \, \overline{\psi}_{i m \sigma}({\bf r})
\left[-\frac{1}{2}\nabla^2+v_{\rm R}({\bf r})\right]\psi_{i^\prime m^\prime \sigma}({\bf r}).
\end{eqnarray*}
The on-site ($i=i^\prime$) matrix is the crystal-field  and the $i\ne i^\prime$ contributions are the hopping integrals.
The Coulomb interaction $H_U$ is given by
\begin{eqnarray*}
H_{U}=\frac{1}{2}
\sum_{i i^\prime  j j^{\prime}}
\sum_{\sigma \sigma^\prime}
\sum_{m  m^\prime} \sum_{  p p^\prime}
U_{m p \; m^\prime p^\prime}^{iji^\prime j^\prime}
c^\dagger_{i m \sigma}
c^\dagger_{j p \sigma^\prime}
c^{\phantom{\dagger}}_{j^{\prime} p^{\prime}\sigma^\prime}
c^{\phantom{\dagger}}_{i^{\prime} m^{\prime}\sigma}.
\end{eqnarray*}
The {\em bare} Coulomb integrals are  
\begin{eqnarray}\label{bare}
U_{n  p \; n^\prime p^\prime }^{iji^\prime j^\prime}&=& \! \int \! d{\bf r}_1 \! \!
    \int  \! d{\bf r}_2 \; 
\overline{\psi}_{in \sigma}({\bf r}_1)
\overline{\psi}_{j p\sigma^\prime}({\bf r}_2)
                 \frac{1}{|{\bf r}_1-{\bf r}_2|}  
\psi_{j^{\prime} p^{\prime} \sigma^{\prime}}({\bf r}_2)
\psi_{i^{\prime} n^{\prime}\sigma}({\bf r}_1)
.
\end{eqnarray}
The term ${H}_{\rm DC}$ (double-counting correction) cancels the electron-electron interaction contained in ${H}^{\rm LDA}$ but also explicitly described by $H_{U}$. Although such term is in principle unknown, reasonable approximations
have been developed in the context of the LDA+$U$ approach \cite{ldau}, a method which shares the model building part with LDA+DMFT, and are successfully used also in LDA+DMFT calculations \cite{book2011,book2012}.

\begin{figure}[t]
 \centering
 \rotatebox{0}{\includegraphics[width=1.11\textwidth]{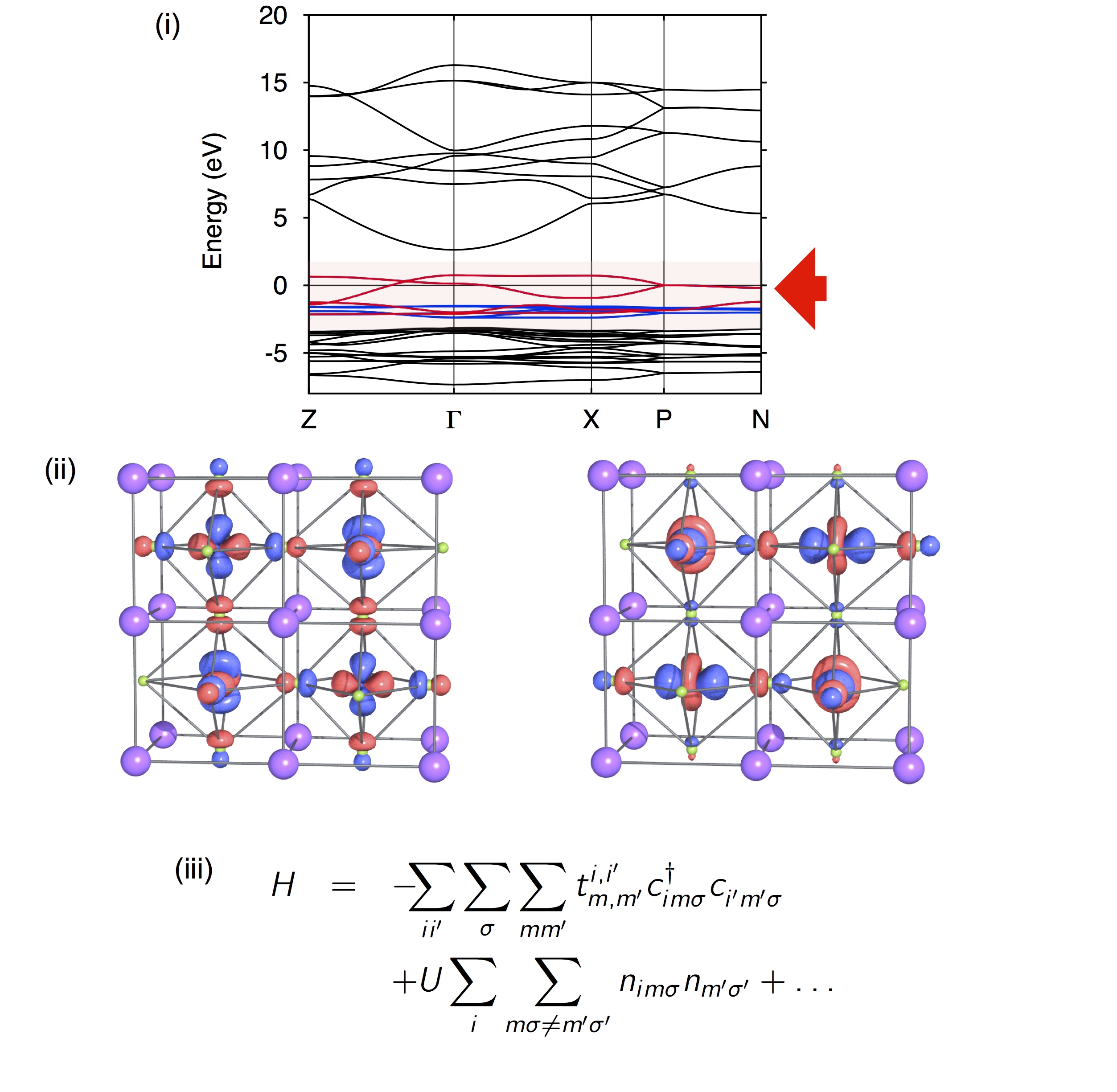}}
 \caption{\label{model1} Building minimal material-specific  many-body models in steps: 
    (i) Full LDA band structure of the perovskite KCuF$_3$: The region with a red background shows the low-energy states,
   the Cu $d$ bands. They split into partially filled $e_g$ (red) and  occupied $t_{2g}$ (blue).
   The $e_g$ bands are identified as the correlated states.
   (ii) Massive downfolding: Crystal-field Wannier functions basis spanning the $e_g$ bands in order (left to right)
        of decreasing crystal-field energy. 
        The Wannier functions are plotted at each Cu site to show the site symmetries.  
  (iii) Low-energy generalized Hubbard model for the $e_g$ bands. }
\end{figure}
The Hamiltonian (\ref{Hele}) still describes the full many-body problem, whose solution  remains inaccessible.
It is therefore necessary to devise approximations to reduce the complexity of the problem to the essential.
The most important one consists in separating the electrons in two types, the {\em correlated} or heavy electrons, for which LDA fails, and the {\em uncorrelated} or {light} electrons, for which LDA works sufficiently well (Fig.~\ref{model1}). 
The correction to LDA, $H_{U}-{H}_{\rm DC}$, is then taken into account explicitly only for the correlated electrons, typically
chosen as those stemming from partially filled localized $d$ and $f$ atomic shells (Fig.~\ref{extension}).
By truncating $H_{U}-{H}_{\rm DC}$ to the heavy-electron sector we implicitly assume that  the main effect of light electrons is the effective {screening} of the Coulomb parameters for heavy electrons. Thus the {\em bare} Coulomb integrals (\ref{bare})
are replaced by {\em screened} parameters. The calculation of effective screened Coulomb couplings remains a major challenge to date. Approximate schemes developed so far are the constrained LDA (cLDA) approach \cite{cLDA} and the constrained RPA (cRPA) method \cite{cRPA}.
In the first the screened $U$ is computed from the second derivative of the total energy as a function of the density;  the hopping integrals between heavy and light electrons are cut to avoid electron transfer between
light and heavy electrons sectors.
In cRPA the polarization (and thus the screened Coulomb interaction) is obtained in the random-phase approximation by downfolding the uncorrelated sector, assuming that the latter is well described by mean field theory.
The last important approximation is the assumption that the Coulomb interaction is local or very fast decaying.
The successes of practical DFT suggest that the long-range Hartree and the mean-field exchange-correlation interaction are well described by the LDA or its extensions; we then can expect that the $H_{U}-{H}_{\rm DC}$ term is all the rest, i.e., that it is local (on-site) or almost local (between first nearest neighbors).

By means of these simplifications we have transformed the full many-body problem (\ref{ele}) to a minimal  material-specific generalized Hubbard model with local or almost local Coulomb interaction. Even such a model cannot be solved exactly; as we will see in the next section, the dynamical mean-field approximation allows us, however, to capture the microscopic mechanisms
behind emergent phenomena such as the Mott transition or orbital order.

In the rest of this paragraph we discuss in some more detail the form of the local Coulomb term.
Correlated electrons  partially retain their atomic character. Thus they are usually identified through the quantum
numbers $lm\sigma$ of the atomic shells from which they stems. If $H_{U}-{H}_{\rm DC}$ is local and correlated electrons belong to a given shell (e.g., $d$ electrons, $l=2$), 
the {\em screened} Coulomb interaction can be written as $H_{U}-{H}_{\rm DC}\sim H_U^l-H_{\rm DC}^l$, with
\begin{eqnarray*}
H_U^l-H_{\rm DC}^l=\frac{1}{2}
\sum_i\sum_{\sigma \sigma^\prime}
\sum_{m  m_{}^\prime} \sum_{\tilde{m}\tilde{m}^\prime   }
U^l_{m \tilde{m} m_{}^\prime \tilde{m}^\prime}
c^\dagger_{im_\sigma}
c^\dagger_{i\tilde{m}\sigma^\prime}
c^{\phantom{\dagger}}_{i\tilde{m}^\prime\sigma^\prime}
c^{\phantom{\dagger}}_{im_{}^\prime\sigma}-H_{\rm DC}^l
\end{eqnarray*}
where $m,m^\prime,\tilde{m},\tilde{m}$ run from $-l$ to $l$, and ${H}_{\rm DC}^l$ is the mean-field (e.g., Hartree or Hartree-Fock) value of $H_U^l$.
The screened Coulomb interaction has the same form of the bare interaction
but it has renormalized  $U^l_{m \tilde{m} m_{}^\prime \tilde{m}^\prime}$ parameters.
For simplicity we discuss it in the basis of
atomic orbitals, $\psi_{nlm}({\bf r})=R_{nl}(r) Y_m^{l}(\theta,\phi)$. 
Thus
\begin{eqnarray*}
U_{m m m^\prime \tilde{m}^\prime}^l
= \sum_{k=0}^{2l} a_{k}^l (m m^\prime, \tilde{m} \tilde{m}^{\prime}) F_{k}^l,
\end{eqnarray*}
where the angular integrals are
\begin{eqnarray*}
a_{k}^l(m m^\prime, \tilde{m} \tilde{m}^{\prime})=\frac{4\pi}{2k+1} \sum_{q=-k}^k 
\langle l m               | Y_q^k | l m^{\prime} \rangle
\langle l \tilde{m} | \overline{Y}_q^k\:| l \tilde{m}^{\prime} \rangle,
\end{eqnarray*}
and the radial Slater integrals are
\begin{eqnarray*}
F_{k}^l=\int\! d{ r}_1 \: r_1^2 \int \! d{r}_2 \: r_2^2 \:  R^2_{nl}(r_1)\frac{r_<^k}{r_>^{k+1}}R_{nl}^2(r_2).
\end{eqnarray*}
The most important Coulomb integrals are the direct
($U_{m m^\prime m m^\prime}^l$) and exchange ($U_{m m^\prime m^\prime m}^l$, with $m\ne m^\prime$)  integrals, which can be expressed as
\begin{eqnarray*}
U_{m m^\prime m m^\prime}^l=U_{m,m^\prime} &=& \sum_{k=0}^{2l} a_k^l(mm,m^{\prime}m^{\prime}) F_k^l,\\
U_{m m^\prime m^\prime m}^l=J_{m,m^\prime} &=& \sum_{k=0}^{2l} a_k^l(mm^\prime, m^{\prime}m) F_k^l.
\end{eqnarray*}
The average Coulomb parameters  are 
\begin{eqnarray*}
U_{avg}&=&\frac{1}{(2l+1)^2}\sum_{m,m^\prime} U_{m,m^\prime} =F_0^2, \\
U_{avg}-J_{avg}&=&\frac{1}{2l(2l+1)}\sum_{m , m^\prime} (U_{m,m^\prime}-J_{m,m^\prime}).
\end{eqnarray*}
For atomic states $U_{\rm avg}$ is very large (typically $15-20$~eV for $d$ electrons) but is drastically reduced by 
screening effects. For $d$ shells ($l=2$) only $F_0^2$, $F_2^2$ and $F_4^2$ contribute to the Coulomb integrals, and  $J_{\rm avg}=(F_2^2+F_4^2)/14$.
For hydrogen-like 3$d$ orbitals, $F_4^2/F_2^2=15/23$, while for realistic 3$d$ orbitals this ratio is slightly smaller; a typical value of the ratio $F_4^2/F_2^2$ is $\sim 0.625=5/8$. 

It is useful to re-express the parameters $U_{m,m^\prime}$ and  $J_{m,m^\prime}$ as a function of 
the following three parameters
\begin{eqnarray*}
U_0&=&     U_{\rm avg} +\frac{8}{7} J_{\rm avg}= U_{\rm avg} +\frac{8}{5} {\cal{J}}_{\rm avg} \\
{\cal{J}}_{avg}&=&\frac{1}{2l(2l+1)}\sum_{m \ne m^\prime} J_{m,m^\prime}=\frac{5}{7} J_{\rm avg} \\
\Delta {\cal{J}}_{avg}&=& {\cal{J}}_{avg}\left(\frac{1}{5}-\frac{1}{9}\frac{F_4^2}{F_2^2} \right) /\left( 1+\frac{F_4^2}{F_2^2}\right)\\
\end{eqnarray*}
Here $U_0$ is the orbital-diagonal direct Coulomb integral; ${\cal{J}}_{avg}$ the average interaction in
the basis of cubic harmonics, typically used in electronic calculations; $\Delta {\cal{J}}_{avg}$ measures the orbital anisotropy of the Coulomb interaction.
For the $d$ shell the matrix $U_{m,m^\prime}-U_0$ may then be written as (see Appendix)
\begin{equation}\label{umm}
\begin{array}{l|cccccc} 
U_{m,m^\prime}-U_0\quad&|xy\rangle& |yz\rangle& |3z^2-r^2\rangle & |xz\rangle & |x^2-y^2\rangle  \\[1ex] \hline\\
|xy\rangle      &0 &-2J_1  &-2J_2 &-2J_1 &-2J_3 \\
|yz\rangle      & -2J_1 &  & -2J_4&-2J_1 &-2J_1 \\
|3z^2-r^2\rangle  & -2J_2 &-2J_4  &0 & -2J_4& -2J_2 \\
|xz\rangle      & -2J_1 &-2J_1  & -2J_4
 & 0 & -2J_1\\
|x^2-y^2\rangle & -2J_3& -2J_1 & -2J_2 & -2J_1 & 0 \\[1ex]
\end{array}
\end{equation}
where the exchange integrals $J_{m,m^\prime}$ appearing in (\ref{umm}) are
$$J_1=    {\cal{J}}_{\rm avg} + \Delta{\cal{J}}_{\rm avg},$$
$$J_2=                 {\cal{J}}_{\rm avg} +3 \Delta{\cal{J}}_{\rm avg},$$ 
$$J_3=     {\cal{J}}_{\rm avg} -5 \Delta{\cal{J}}_{\rm avg},$$ 
$$J_4=    {\cal{J}}_{\rm avg} -3 \Delta{\cal{J}}_{\rm avg}.$$
The Coulomb anisotropy $\Delta{\cal{J}}_{\rm avg}$ is crucial for a proper description of the multiplet structure, and is particular important for systems 
close to spin-state transitions, such as cobaltates \cite{ss}.
Various approximations of the Coulomb interaction  are often adopted in LDA+DMFT calculations; they are typically introduced to  reduce the complexity of the calculation and the CPU-time or to make the problem tractable in the first place. One has to keep in mind that these approximations alter the structure of the multiplets, and their validity has to be considered case by case. 
Perhaps the most common approximation of the Coulomb interaction is the density-density  approximation, in which only Coulomb terms that can be expressed as density-density interaction are retained. This approximation is typically adopted when the Hirsch-Fye quantum Monte Carlo (HF-QMC) algorithm \cite{hirsch} is used as DMFT quantum-impurity solver. To go beyond it with QMC has required the development of a new algorithm, the continuous-time QMC approach (CT-QMC) \cite{ctqmc}.

\section{The Dynamical Mean-Field Approximation}
The separation of electrons in light and heavy greatly simplifies the problem, 
reducing the many-body Hamiltonian (\ref{ele}) to a generalized Hubbard model.
Nevertheless, the exact solution of such a many-body problem remains out of reach. The DMFT \cite{dmft} is to date the best
approximate method which still retains the essential ingredients to explain strong correlation phenomena
such as the Mott metal-insulator transition or orbital ordering.

\begin{figure}[t]
\centering
\rotatebox{0}{\includegraphics[width=1.0325\textwidth]{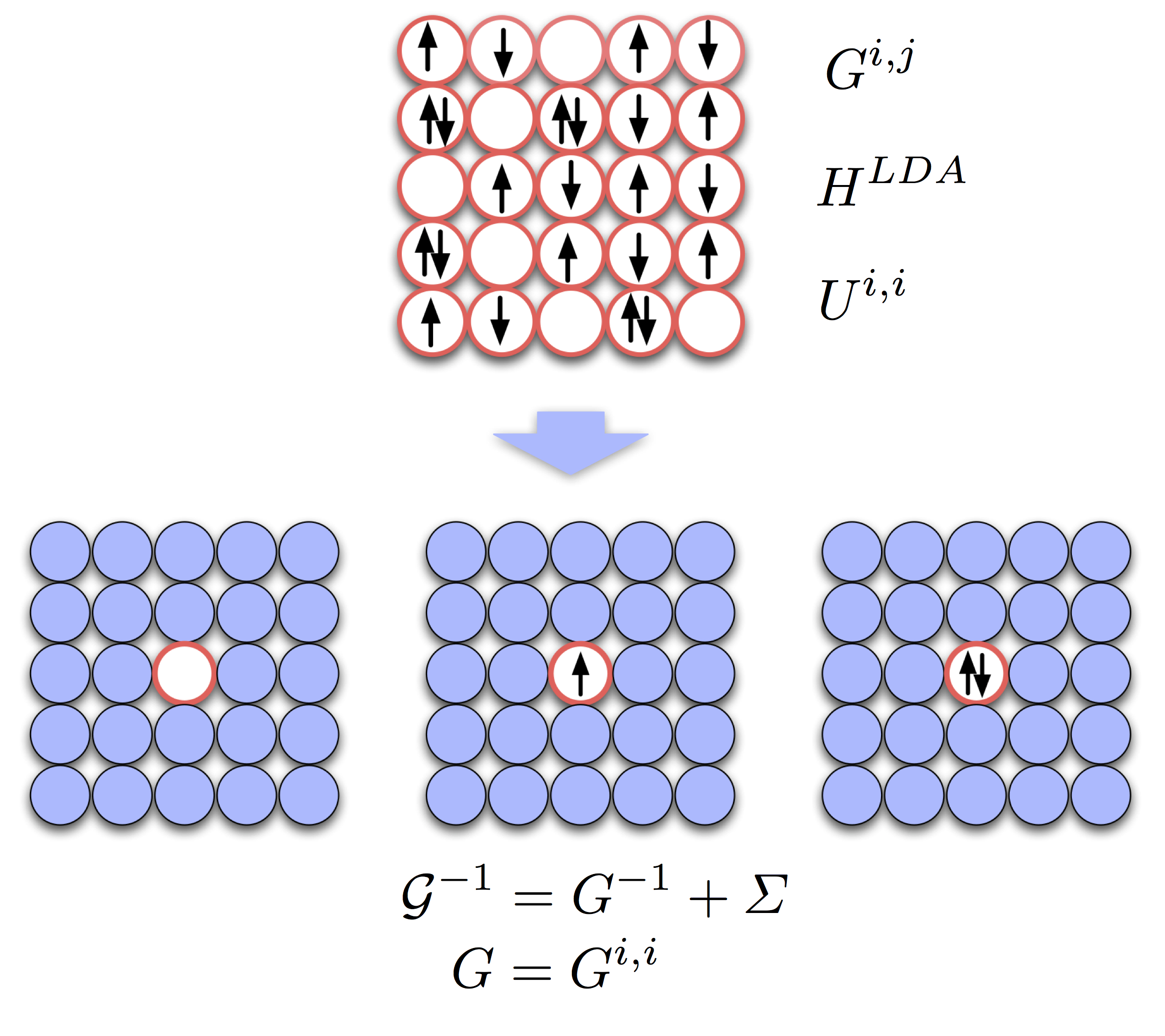}}
\caption{\label{map} Schematic illustration of the dynamical mean-field approximation. The lattice Hubbard model is mapped onto an effective quantum-impurity problem
satisfying the condition $G=G^{i,i}$, where $G^{i,i}$ is the local lattice Green-function matrix
and $G$ the Green-function matrix of the impurity problem.
The self-energy matrix is dynamical ($\omega$-dependent) but local ({\bf k}-independent).
The dynamical mean-field theory is exact in the limit of infinite coordination number \cite{dmft}.}
 \end{figure}

In DMFT the lattice many-body Hubbard model is mapped  onto an effective single-impurity model
 which describes a single correlated site in an effective bath  (Fig.~\ref{map}).
The exact solution of a such a quantum-impurity problem is hard, but this time various 
numerically exact techniques are available.

\begin{figure}[t]
\centering
\rotatebox{0}{\includegraphics[width=1.01\textwidth]{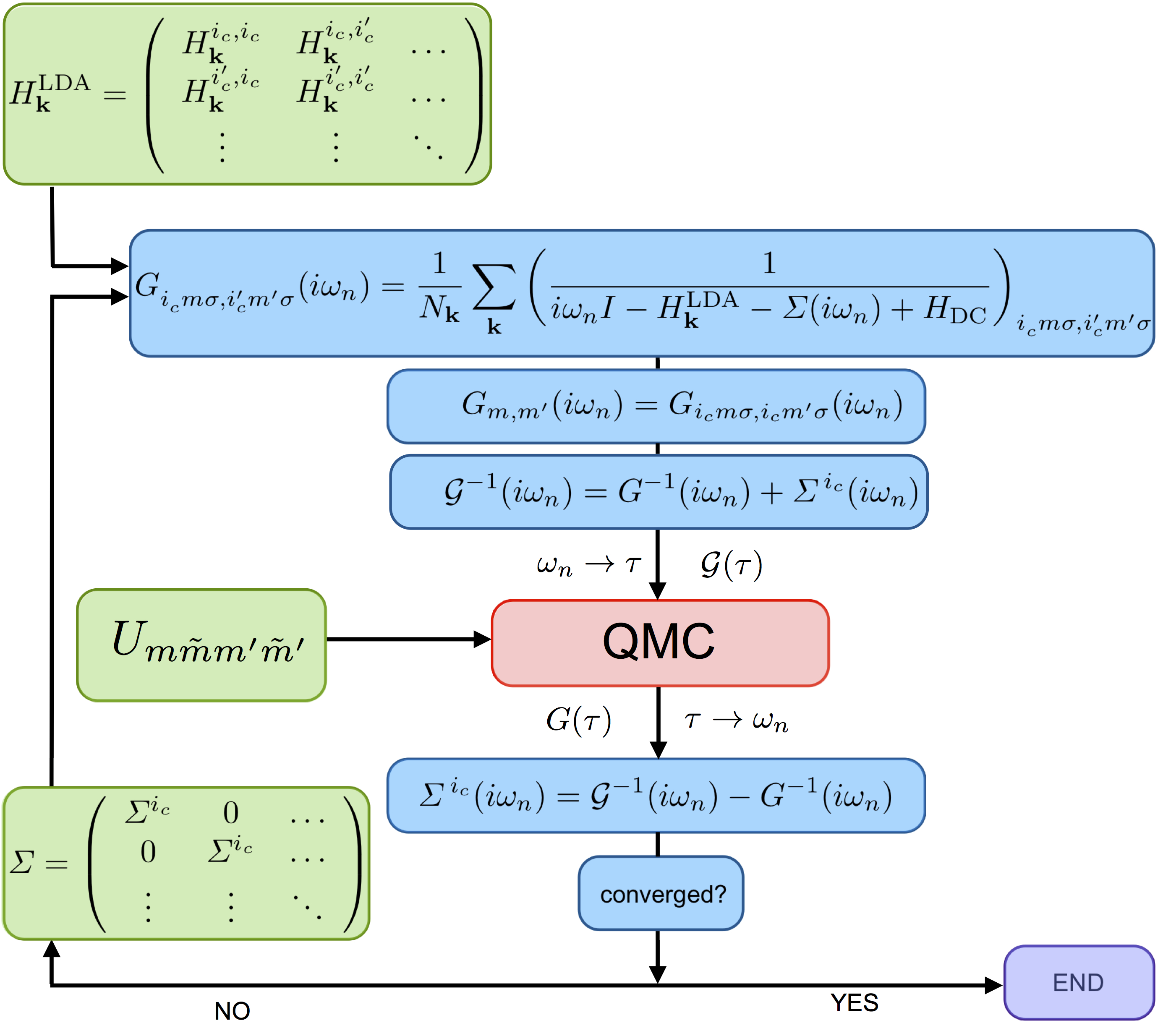}}
\caption{\label{self} LDA+DMFT self-consistency loop for a system with $\{ m \}$ orbital degrees of freedom and at least two equivalent correlated sites. We assume that a quantum Monte Carlo impurity solver is used, and that the quantum-impurity problem is then solved in imaginary time $\tau$ and Matsubara frequencies $\omega_n$.}
 \end{figure}

The various steps of LDA+DMFT are shown in Fig.~\ref{self}.
Let us assume we followed the procedure described in the previous section.
In the example of the KCuF$_3$ perovskite of Fig.~\ref{model1} we identify  the $e_g$ states as correlated electrons.
The minimal model for such a system is a 2-band $e_g$ Hubbard model, and the smallest
possible unit cell has two equivalent correlated sites.
More generally, we can assume that a system is
described by a unit cell with $i_c=1,\dots, n_c$ equivalent correlated sites and at each correlated site
we label  the correlated orbitals with $\{m\sigma \}$; a number of non-correlated orbitals and sites are 
also in general included in the model. 

To solve with DMFT the Hubbard-like model for our system we first map it onto a quantum-impurity
problem. Next we solve the latter self-consistently, i.e.,
with the constraint that the impurity Green-function matrix  $G(\omega)$ 
equals the local lattice Green-function matrix $G^{\: i_c,i_c}(\omega),$ 
$$
G_{m,m^\prime}( \omega )=G^{\: i_c,i_c}_{m\sigma, m^\prime\sigma}( \omega).
$$
To do this we use an iterative procedure.
First we calculate the local lattice Green-function matrix
for a given bath, i.e.,

\begin{eqnarray*}
G^{\: i_c,i_c}_{m\sigma, m^\prime\sigma}(\omega)=\frac{1}{N_{\bf k}}  \sum_{\bf k} 
\left(\frac{1}{{(\omega+\mu )I-{H}_{\bf k}^{\rm LDA}-\Sigma(\omega) +{H}_{\rm DC}}}\right)_{i^{\phantom{'}}_c m \sigma,i_c^\prime m^\prime\sigma},
\end{eqnarray*}
%
where ${H}_{\bf k}^{\rm LDA}$ is the LDA Hamiltonian in {\bf k} space, and $\Sigma (\omega)$ is
the self-energy matrix; to first iteration we can assume that the bath Green-function matrix
is the local LDA Green-function matrix and that the self-energy is therefore zero.
In the next iterations, the DMFT self-energy matrix is non-zero in the correlated sector
(sites $\{i_c\}$ and orbitals $\{m\}$).
Furthermore it is local, i.e.,
$$
\Sigma_{ m\sigma, m^\prime\sigma}^{i_c^{\phantom{'}},i_c^\prime}(\omega)=
\delta_{i_c^{\phantom{'}},i_c^\prime}\Sigma^{\,i_c \sigma}_{m,m^\prime}(\omega).
$$
For two equivalent correlated sites $i_c^{\phantom{'}}$ and $i_c^\prime$, space group symmetries transform $\Sigma^{\,i_c^{\phantom{'}} \sigma}_{m, m^\prime}$ into $\Sigma^{\,i_c^\prime \sigma}_{m, m^\prime}$. For convenience we can instead use the same symmetries to transform the LDA Hamiltonian  in such way that its on-site blocks are identical for equivalent correlated sites. 
Then 
$$
\Sigma^{\,i_c \sigma}_{m\sigma, m^\prime\sigma} (\omega)
=\Sigma^{\sigma}_{m,m^\prime}(\omega).
$$
In the paramagnetic phase an additional relation holds, i.e. 
$$
\Sigma^{\sigma}_{m, m^\prime}(\omega)
=\Sigma^{ -\sigma}_{m, m^\prime}=\Sigma_{m, m^\prime}(\omega).
$$ 
The bath Green function of the single impurity model can be obtained from the Dyson equation
\begin{eqnarray*} 
\quad
{\cal{G}}^{-1}(\omega)=G^{-1} (\omega)+\Sigma(\omega).
\end{eqnarray*}
\begin{figure}[t]
\centering
\rotatebox{0}{\includegraphics[width=0.93\textwidth]{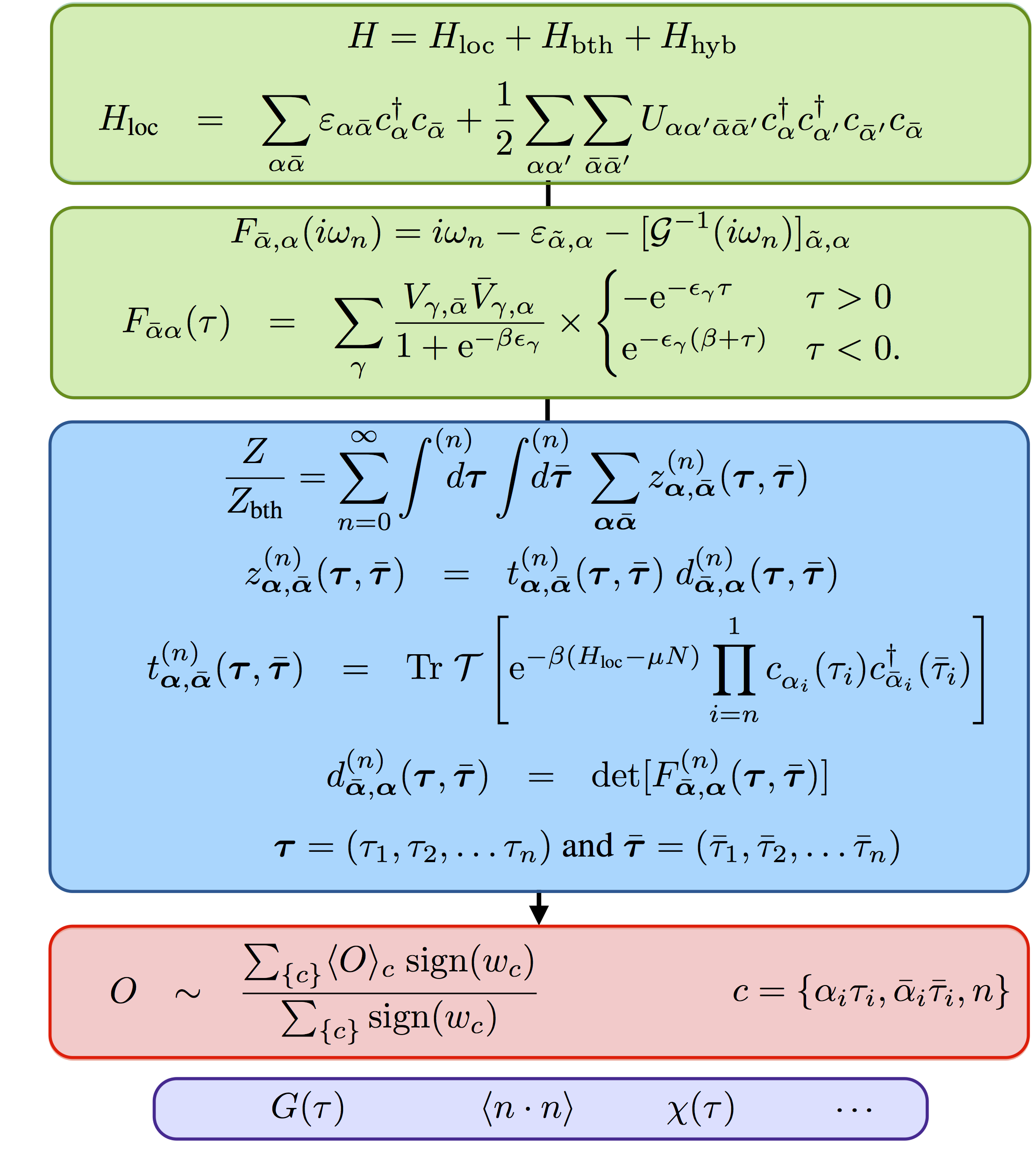}}
\caption{\label{cthyb}Schematic representation of the most important steps in CT-HYB QMC \cite{ctqmc},
following the notation of Ref.~\cite{ctqmcflesch}.
The impurity Hamiltonian is split into local ($H_{\rm loc}$), bath ($H_{\rm bth}$)
and hybridization ($H_{\rm hyb}$) term. 
The bath Green function yields the hybridization function $F$, which together with the local
Hamiltonian makes up the input to the QMC simulation. 
In the next step, the partition function is
expanded in even orders $2n$ of $H_{\rm hyb}$ and bath degrees of freedom are integrated out. To obtain a given observable,
configurations (made of expansion orders $n$, flavors ${\alpha_i}=m_i\sigma_i $, and imaginary times ${\bf \tau}_i,\bar\tau_i$) are sampled
via a Monte Carlo procedure. The bottleneck is the calculation
of the local trace $t^{(n)}$ which requires multiplications of creator and annihilator matrices
as well as the propagation of vectors in imaginary time. In the so-called Krylov scheme
a number basis is used so that all operators are sparse matrices and the Lanczos algorithm
is used in the propagation of vectors. A considerable reduction in computational
time can be achieved by exploiting symmetries. }
 \end{figure}
The quantum-impurity problem, defined by the correlated impurity and the bath Green-function matrix, is then solved via a quantum-impurity solver, which yields a new impurity Green-function matrix, $G_{m, m^\prime} (\omega)$.
A new self-energy is obtained from the Dyson equation
\begin{eqnarray*} 
\quad
\Sigma(\omega)={\cal{G}}^{-1}(\omega)-G^{-1} (\omega).
\end{eqnarray*}
The procedure is repeated till self-consistency  is reached.

The core of the DMFT simulation is the solution of the multi-orbital quantum-impurity model, i.e. finding
$G(\omega)$ for a given $\cal{G}(\omega)$ and Coulomb interaction. Several numerically exact approaches exist, the most important being Quantum Monte Carlo \cite{hirsch,ctqmc}, Lanczos \cite{lanczos,erik}, the numerical renormalization group, and many more. 
Each technique is best suited to deal with certain types of problems. Perhaps  to date the most flexible methods 
are those based on QMC techniques.
The Hirsch-Fye QMC \cite{hirsch} approach is very general. A limitation
is that spin-flip and pair-hopping terms have to be neglected; in all cases in which the multiplet structure is not crucial, this is, however, a very good approximation. A second limitation of HF-QMC is that, for multi-orbital systems, low temperatures can only be reached at prohibitive cost. The continuous-time QMC technique \cite{ctqmc} is based on the 
expansion of the partition function in powers of the hybridization (CT-HYB) or 
the interaction (CT-INT), and can treat spin-flip and pair-hopping terms; 
the CT-HYB algorithm is the one best suited to reach experimental temperatures
for realistic models; CT-INT can deal more  efficiently with clusters extensions of DMFT. 
A schematic representation of the CT-HYB technique is given in Fig.~\ref{cthyb}.
A drawback of all QMC techniques discussed so
far is that they yield results on the imaginary axis, and the analytic continuation requires techniques
such as the maximum entropy method \cite{Jarrell} or stochastic approaches \cite{Mit}. Real-frequency methods such as exact diagonalization or Lanczos have the advantage
that they give access directly to the real axis, and are fast; they require however large memory;
furthermore, in these techniques the bath is discretized, and the convergence with the number of bath
sites can be very slow, in particular in small-gap systems.
Apart from numerically exact approaches, approximate solvers are also often adopted: various Hubbard approximations, the iterative perturbation theory,  the one-crossing approximation, the Gutzwiller variational approach, or others.

\begin{figure}[t]
\centering
\rotatebox{0}{\includegraphics[width=1.0\textwidth]{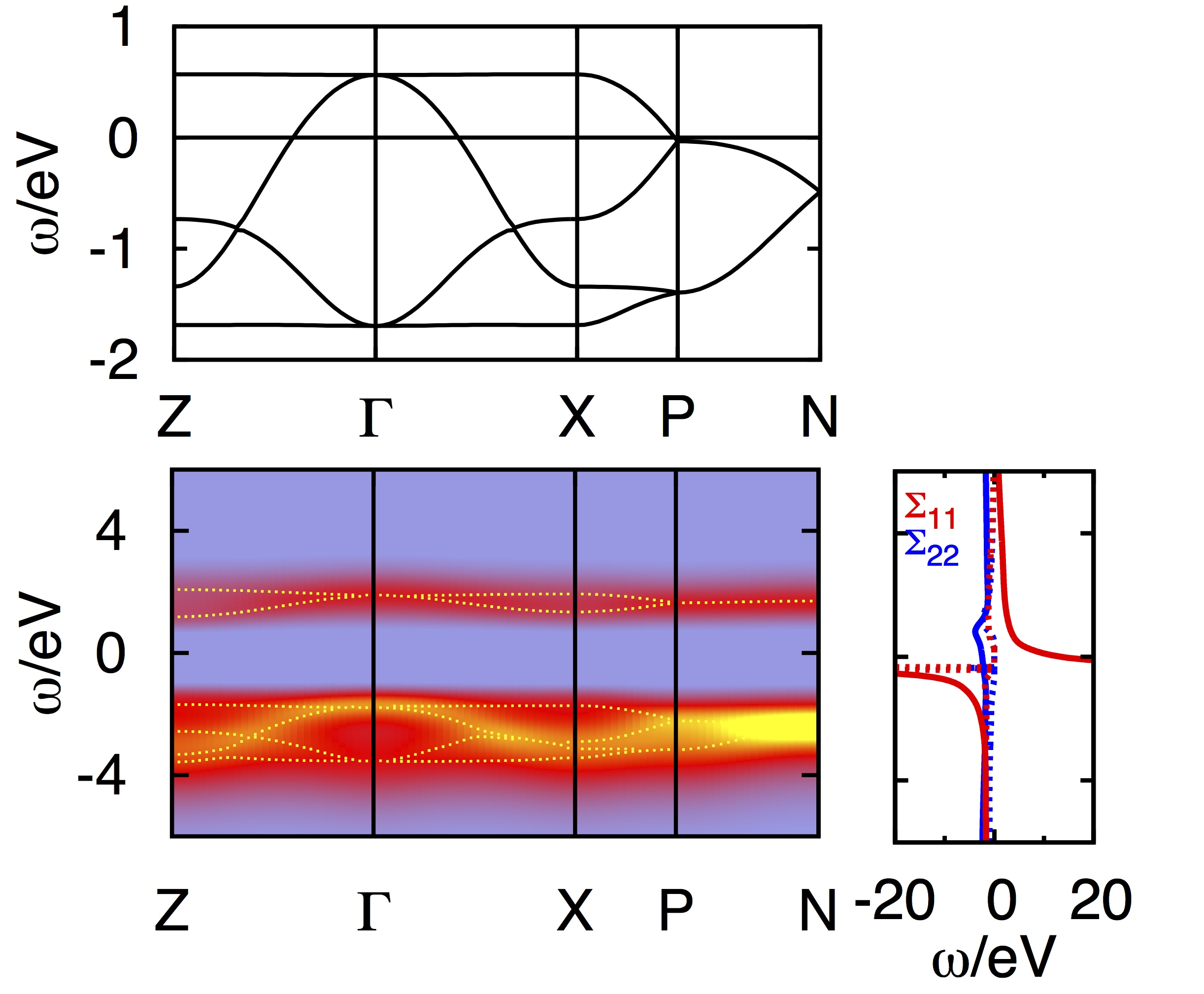}}
\caption{\label{bands}LDA (top) and corresponding LDA+DMFT (bottom) $e_g$ band structure of the Mott insulator KCuF$_3$ in the 
orbitally ordered phase. Calculations are for the ideal cubic structure, massive downfolding to the $e_g$ bands, 
 and for a temperature below 
the super-exchange orbital-order critical temperature $T_{\rm KK}$. The real part of the
self-energy of the hole orbital ($\Sigma_{11}$) diverges in the gap leading to a Mott state.
Adapted from Refs.~\cite{book2011,oo1}.}
 \end{figure}
We have up to now not specified the actual size of the LDA Hamiltonian. 
Wannier functions allow for {\em massive downfolding}, i.e., for reducing the Hamiltonian to the minimal
basis set describing the correlated electrons. In the example of KCuF$_3$, considered in the previous section,
the smallest possible basis is that made of $e_g$ states only. The $e_g$ Wannier functions spanning the $e_g$ bands
are longer ranged
than atomic orbitals, because they carry the information on the lattice and the bonding.
This can be seen in the orbitals in Fig.~\ref{model1}. The LDA+DMFT band structure for the $e_g$ bands
is shown in Fig.~\ref{bands}.
The advantage of massive downfolding is that the double-counting correction can be incorporated in the chemical potential and does not need to be calculated explicitly. 
In some cases, however, one has to include  non-correlated electrons in the calculation.
This is the case if, for example, $p$-$d$ charge-transfer effects play an important role.
In these cases a larger basis set is used and the double-counting
correction has to be calculated explicitly. 

The extension of the LDA+DMFT scheme discussed above to the spin-polarized case (e.g., for studying ferro- or antiferro-magnetic phases) is straightforward, provided that the right unit cells and symmetries are used.
The LDA+DMFT scheme can be also easily extended to clusters, e.g., by using a supercell as quantum impurity;  other non-local extensions of DMFT 
are the dynamical-cluster approximation (DCA) \cite{dca}, the dual-fermion approach, or the GW+DMFT approach
(see Refs.~\cite{book2011,book2012} ). 
One has to keep in mind that increasing the number of sites and degrees of freedom the problem becomes progressively harder;
in QMC calculations the computational time becomes quickly prohibitively long and the infamous minus sign problem can arise; in Lanczos-based calculations one might need a computer with more GB of memory than there are atoms in the visible universe.
Thus in practice the calculations can reach quickly the feasibility limit,
even with the help of modern  massively-parallel supercomputers.

As final remark, LDA+DMFT calculations can also be performed {\em charge self-consistently}. 
If we assume that LDA describes uncorrelated electrons sufficiently well, the readjustments in the uncorrelated sector can be calculated by making the total charge density and the reference potential consistent within the LDA, however with the constraints provided by the DMFT solution of the Hubbard model.  
This requires to work with the full Hamiltonian and, again, to account explicitly for the double-counting correction.
 
\begin{figure}[t]
\centering
\rotatebox{0}{\includegraphics[width=0.8\textwidth]{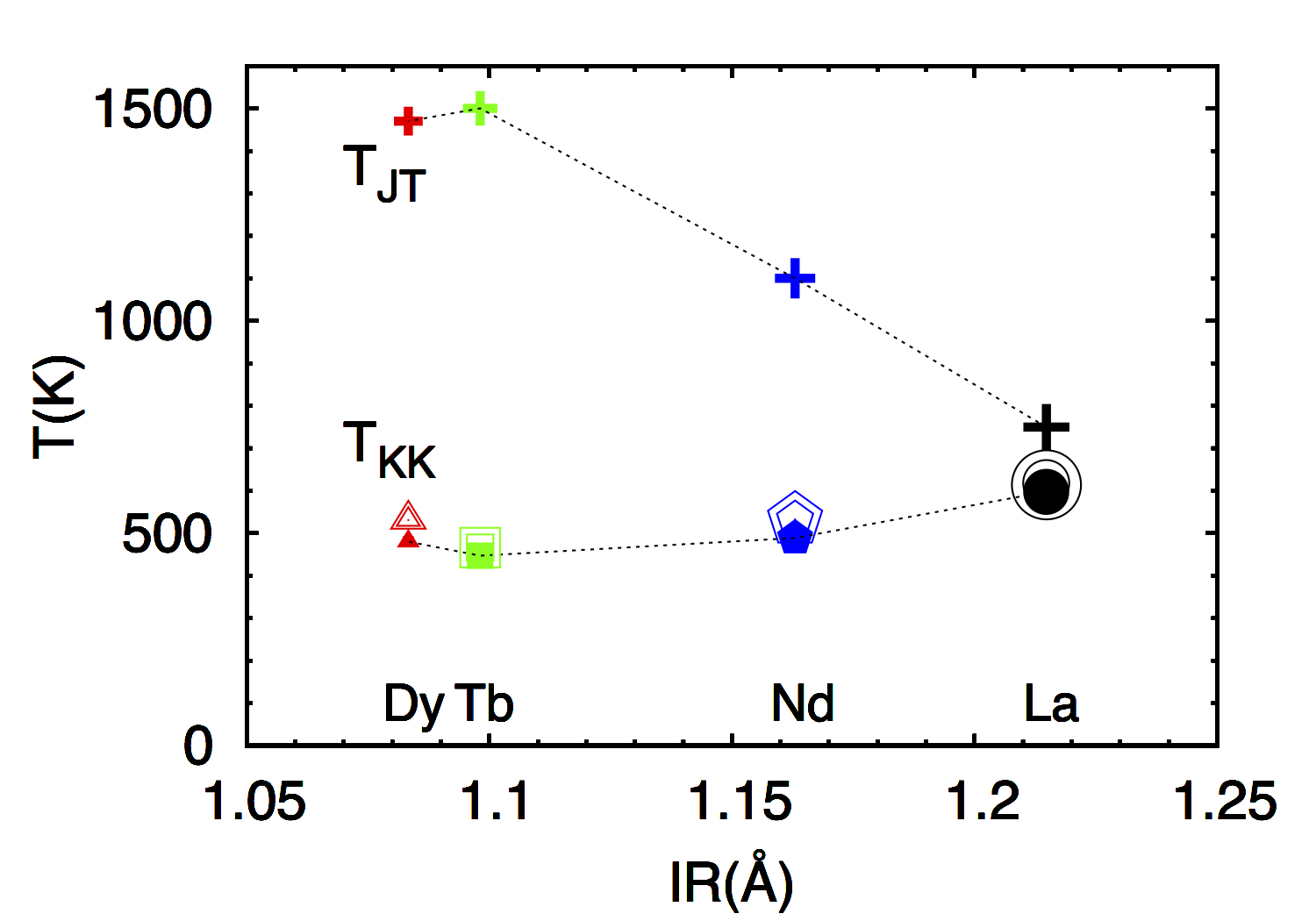}}
\caption{\label{tkk} 
$T_{\rm KK}$, the critical temperature for orbital order due to super-exchange only, versus
the experimental orbital melting temperature, $T_{\rm JT}$. 
Calculations were done with density-density Coulomb interaction (full symbols: HF-QMC; small open symbols: CT-HYB)
and full Coulomb interaction (larger open symbols). The results show
that spin-flip and pair-hopping terms do not affect $T_{\rm KK}$ in a sizable way.
Adapted from Refs.~\cite{oo2,oo3,ctqmcflesch}.}
 \end{figure}
\section{The Origin of Orbital Order}

In this paragraph we illustrate a paradigmatic application of the LDA+DMFT technique.
Orbital order phenomena play a crucial role in the physics of strongly correlated
oxides, such as the $e_g$ systems KCuF$_3$ and LaMnO$_3$.
The origin of orbital order has been therefore debated since long.
Two mechanisms have been proposed. 
The first is electron-phonon coupling \cite{ep}, which naturally yields a static co-operative Jahn-Teller
distortion. The order arises from the splitting of the partially filled degenerate $e_g$ levels 
due to the Jahn-Teller crystal-field. LDA+DMFT calculations have shown that even a small crystal field is sufficient, 
because it is enhanced by Coulomb repulsion effects; this phenomenon has been observed
in very diverse materials \cite{d1,d2,ruthenates}.
The second mechanism is the purely electronic Kugel-Khomskii super-exchange \cite{kk}.
In this picture the static co-operative Jahn-Teller distortion is a consequence, rather than the cause, of orbital order.
In real systems it is very difficult to disentangle the two effects, because both
lead to similar co-operative distortions. Total energy studies based on
antiferromagnetic LDA+$U$  \cite{ldau,ku} and more recently paramagnetic LDA+DMFT \cite{leonov}
calculations show that, in order  to explain the presence of
the Jahn-Teller co-operative distortion, we have to take into account
the Coulomb interaction. However, in both KCuF$_3$ and LaMnO$_3$ the magnetic transition
temperature ($\sim$40~K for KCuF$_3$ and $\sim$140~K for LaMnO$_3$) is sizably smaller than the orbital-ordering temperature, suggesting that quite different mechanisms are involved in the two phenomena.

Recently we devised a procedure which allows us to disentangle the super-exchange from the electron-phonon mechanism \cite{oo1}.
It consists in performing LDA+DMFT calculations of the orbital polarization versus temperature 
for a series of progressively less-distorted materials
and determining in this way the transition-temperature due to super-exchange {\em only}.
Using this procedure for KCuF$_3$  we find  $T_{\rm KK}\sim 350~$K \cite{oo1}.
In this system $T_{\rm JT}$, the experimental 
temperature at which the co-operative Jahn-Teller distortion disappears in X-ray or neutron scattering data,
is close or perhaps above the melting temperature \cite{ghigna}, i.e.,
$T_{\rm KK} $ is sizable but much smaller than  $T_{\rm JT}$. 
This shows that super-exchange is large  but not sufficient to determine the presence of a co-operative
Jahn-Teller distortion at high temperatures. 

The situation is much more complex for the rare-earth manganites.
Figure~\ref{tkk} shows $T_{\rm KK}$ for the $R$MnO$_3$
 series \cite{oo2,oo3,ctqmcflesch}, compared to $T_{\rm JT}$.
Remarkably, for LaMnO$_3$ $T_{\rm KK}$ is almost identical to $T_{\rm JT}$.
However, $T_{\rm JT}$ has been identified as the temperature at which an orbital order-to-disorder transition  
occurs \cite{melting}. The JT distortions have be reported to survive in nano-clusters up to 1150 K \cite{clusters}.
Furthermore, the coincidence $T_{\rm KK}\sim T_{\rm JT}$ only occurs for LaMnO$_3$.
In the rest of the series $T_{\rm KK}$ remains more or less the same, while the experimental
$T_{\rm JT}$ becomes as large a 1500 K (Fig.~\ref{tkk}).
This strongly suggests that super-exchange does not determine the order-to-disorder transition.
Finally, we find that tetragonal crystal-field splitting arising from other distortions further 
reduce the effective $T_{\rm KK}$ \cite{oo3}.
In conclusion, our results show that in all considered materials, super-exchange effects, although very large, play a small role in the observed melting of orbital order.

\section{Conclusions and Outlook}
The LDA+DMFT method has opened new horizons for first-principles calculations for strongly correlated
system.  In the last decades it has proven an extremely successful technique for unraveling the physics of a large variety
of strongly
correlated materials.  It is difficult even to list all its successes so far. A partial overview can be found
in reviews \cite{georges,georges2,imada,kotliar}, as well as in \cite{book2011}.
Along the years we have learned that {\em details do matter}, for example crystal-fields an order of magnitude 
smaller than the band width can favor the Mott transition or trigger orbital order \cite{d1,d2,ruthenates,oo1}.

Electronic structure codes based on LDA+DMFT are slowly becoming available in combination 
with most popular DFT codes. Modern parallel supercomputers and algorithmic developments are 
making it possible to solve with DMFT always more complex quantum-impurity models. 
We can nowadays perform charge-self consistent calculations,  optimize structures,
 calculate phonon spectra and response functions.
It is not difficult to imagine that in the not too far future LDA+DMFT codes will become as complex, systematic
and general as modern DFT codes.

Important challenges are however still ahead. 
Non-local effects remain very difficult to describe efficiently.
Screening effects are mostly accounted for only at the cLDA or cRPA level. 
For double-counting corrections only practical recipes exist.    
To further increase the complexity of the quantum-impurity problem 
(i.e. to make it more realistic) it is likely that new efficient quantum impurity solvers 
or even entirely new ideas have to be developed.
Perhaps the most important challenge is to identify which details are important
for specific classes of problems and systems and to work on the development of optimal
schemes to take them all into account.

\section*{Acknowledgment}
Support of the Deutsche Forschungsgemeinschaft through FOR1346 is gratefully acknowledged.  

\clearpage
\section*{Appendix}

\subsection*{Gaunt coefficients and Coulomb integrals}

The two-index Coulomb integrals can be written as
\begin{eqnarray*}
U_{m,m^\prime} &=& \sum_{k=0}^{2l} a_k^l(mm,m^{\prime}m^{\prime}) F_k=\sum_{k=0}^{2l} b_k^l(m,m^{\prime}) F_k^l,\\
J_{m,m^\prime} &=& \sum_{k=0}^{2l} a_k^l(mm^\prime, m^{\prime}m) F_k=\sum_{k=0}^{2l} c_k^l(m,m^{\prime}) F_k^l,
\end{eqnarray*}
where
\begin{eqnarray*}
a_k^l(m_{\alpha}^{\phantom{\prime}} m^\prime_{\alpha}, m_\beta^{\phantom{\prime}} m_\beta^\prime)
=\frac{4\pi}{2k+1}\sum_{q=-k}^k 
\langle lm_\alpha^{\phantom{\prime}}|Y^k_q|lm_{\alpha}^\prime\rangle
\langle lm_\beta^{\phantom{\prime}}|\overline{Y}^k_q|lm_{\beta}^\prime\rangle.
\end{eqnarray*}

In the basis $\{ m\}$ of spherical harmonics, the coefficients  $G_k(m,m^\prime)=\langle lm|Y^k_q|lm'\rangle$ with $k=2,4$ are given by
\begin{displaymath}
\begin{array}{@{\hspace{3ex}}c@{\hspace{6ex}}c@{\hspace{3ex}}c@{\hspace{6ex}}c}
  \displaystyle G_2=\frac{1}{7\sqrt{4\pi}}
\left[\!\begin{array}{rrrrr}
 -\sqrt{20} & \sqrt{30} &  -\sqrt{20} &         0  &         0\\
  -\sqrt{30} & \sqrt{ 5} & \sqrt{ 5} &  -\sqrt{30} &         0\\
  -\sqrt{20} & -\sqrt{ 5} & \sqrt{20} &  -\sqrt{ 5} &  -\sqrt{20}\\
         0  & -\sqrt{30} & \sqrt{ 5} & \sqrt{ 5} &  -\sqrt{30}\\
         0  &         0  &  -\sqrt{20} & \sqrt{30} &  -\sqrt{20}
\end{array}\!\right]\\[8ex]
     \displaystyle G_4=\frac{1}{7\sqrt{4\pi}}
\left[\!\begin{array}{rrrrr}
         1  &  -\sqrt{ 5} & \sqrt{15} & -\sqrt{35} & \sqrt{70}\\
\sqrt{ 5} &         -4  & \sqrt{30} & -\sqrt{40} & \sqrt{35}\\
 \sqrt{15} &  -\sqrt{30} &         6  &  -\sqrt{30} & \sqrt{15}\\
 \sqrt{35} &  -\sqrt{40} & \sqrt{30} &         - 4  & \sqrt{ 5}\\
 \sqrt{70} &  -\sqrt{35} & \sqrt{15} &  -\sqrt{ 5} &         1
\end{array}\!\right].
\end{array}
\end{displaymath}

Thus for $l=2$ we have
\begin{displaymath}
\begin{array}{l@{\hspace{5ex}}c@{\hspace{5ex}}c@{\hspace{5ex}}}
\displaystyle b_0^2\!=\!\!
\left[\begin {array}{r@{\hspace{1ex}}r@{\hspace{1ex}}r@{\hspace{1ex}}r@{\hspace{1ex}}r@{\hspace{1ex}}}
                              1&1&1& 1  &1\\
                            1& 1& 1&1 & 1\\
                            1& 1& 1&1 & 1\\
                            1& 1& 1&1 & 1\\
                            1& 1& 1&1 & 1
\end{array} \right] 
\hspace*{4ex}\displaystyle b_2^2\!=\!\frac{1}{49}\!
\left[\begin {array}{r@{\hspace{1ex}}r@{\hspace{1ex}}r@{\hspace{1ex}}r@{\hspace{1ex}}r@{\hspace{1ex}}}
                           4&-2&-4&-2& 4\\
                            -2& 1& 2&1 &-2\\
                            -4& 2& 4&2 &-4\\
                            -2& 1& 2&1 &-2\\
                             4&-2&-4&-2& 4
\end{array} \right] \\ \\ 
\hspace*{4ex} \displaystyle b_4^2\!=\!\frac{1}{49}\frac{1}{9}\!
\left[ \begin {array}{r@{\hspace{1ex}}r@{\hspace{1ex}}r@{\hspace{1ex}}r@{\hspace{1ex}}r@{\hspace{1ex}}}
1&-4&6&-4&1\\ 
-4&16&-24&16&-4\\ 
6&-24&36&-24&6\\ 
-4&16&-24&16&-4\\ 
1&-4&6&-4&1
\end {array} \right] \\
\end{array}
\end{displaymath}

\begin{displaymath}
\begin{array}{c@{\hspace{5ex}}c@{\hspace{5ex}}c@{\hspace{5ex}}}
\displaystyle c_0^2\!=\!\!
\left[\begin {array}{r@{\hspace{1ex}}r@{\hspace{1ex}}r@{\hspace{1ex}}r@{\hspace{1ex}}r@{\hspace{1ex}}}
                            1&0&0& 0  &0\\
                            0& 1& 0&0 & 0\\
                            0& 0& 1&0 & 0\\
                            0& 0& 0&1 & 0\\
                            0& 0& 0&0 & 1
\end{array} \right] 
\hspace*{4ex} \displaystyle c_2^2\!=\!\frac{1}{49}\!
\left[ \begin {array}{r@{\hspace{1ex}}r@{\hspace{1ex}}r@{\hspace{1ex}}r@{\hspace{1ex}}r@{\hspace{1ex}}}
                              4&6&4&0&0\\ 
                              6&1&1&6&0\\ 
                              4&1&4&1&4\\
                              0&6&1&1&6\\ 
                              0&0&4&6&4
\end {array}  \right] 
\hspace*{4ex} \displaystyle c_4^2\!=\!\frac{1}{49}\frac{1}{9}\!
\left[ \begin {array}{r@{\hspace{1ex}}r@{\hspace{1ex}}r@{\hspace{1ex}}r@{\hspace{1ex}}r@{\hspace{1ex}}}
                             1&5&15&35&70\\
                             5&16&30&40&35\\ 
                            15&30&36&30&15\\ 
                            35&40&30&16&5\\ 
                            70&35&15&5&1
                            \end {array}\right] 
\end{array}
\end{displaymath}\\[.5ex]

Instead, in the basis of real harmonics, the coefficients become
\begin{displaymath}
\begin{array}{l@{\hspace{5ex}}c@{\hspace{5ex}}c@{\hspace{5ex}}}
\displaystyle b_0^2\!=\!\!
\left[\begin {array}{r@{\hspace{1ex}}r@{\hspace{1ex}}r@{\hspace{1ex}}r@{\hspace{1ex}}r@{\hspace{1ex}}}  
                            1& 1& 1&1 &1\\
                            1& 1& 1&1 & 1\\
                            1& 1& 1&1 & 1\\
                            1& 1& 1&1 & 1\\
                            1& 1& 1&1 & 1
\end{array} \right] 
\hspace*{2ex}  \displaystyle b_2^2\!=\!\frac{1}{49}\!
\left[
\begin {array}{r@{\hspace{1ex}}r@{\hspace{1ex}}r@{\hspace{1ex}}r@{\hspace{1ex}}r@{\hspace{1ex}}} 
4&-2&-4&-2&4\\ -2&4&2&-
2&-2\\
-4&2&4&2&-4\\
-2&-2&2&4&-2
\\4&-2&-4&-2&4\end {array}
\right]  \\ \\
\hspace*{2ex} \displaystyle b_4^2\!=\!\frac{1}{49}\frac{1}{9}\!
\left[ 
\begin {array}{r@{\hspace{1ex}}r@{\hspace{1ex}}r@{\hspace{1ex}}r@{\hspace{1ex}}r@{\hspace{1ex}}} 
36&-4&6&-4&-34\\
-4&36&-24&-4&-4\\6&-24&36&-24&6\\
-4&-4&-24&36&-4\\-34&-4&6&-4&36\end {array}
\right] 
\end{array}
\end{displaymath}
\begin{displaymath}
\begin{array}{c@{\hspace{5ex}}c@{\hspace{5ex}}c@{\hspace{5ex}}}
\displaystyle c_0^2\!=\!\!
\left[\begin {array}{r@{\hspace{1ex}}r@{\hspace{1ex}}r@{\hspace{1ex}}r@{\hspace{1ex}}r@{\hspace{1ex}}}  
                            1&0&0& 0  &0\\
                            0& 1& 0&0 & 0\\
                            0& 0& 1&0 & 0\\
                            0& 0& 0&1 & 0\\
                            0& 0& 0&0 & 1 
\end{array} \right] 
\hspace*{4ex}  \displaystyle c_2^2\!=\!\frac{1}{49}\!
\left[ \begin {array}{r@{\hspace{1ex}}r@{\hspace{1ex}}r@{\hspace{1ex}}r@{\hspace{1ex}}r@{\hspace{1ex}}}  
4&3&4&3&0\\
3&4&1&3&3\\
4&1&4&1&4\\
3&3&1&4&3\\
0&3&4&3&4
\end {array}  \right] 
 \hspace*{4ex}\displaystyle c_4^2\!=\!\frac{1}{49}\frac{1}{9}\!
\left[ \begin {array}{r@{\hspace{1ex}}r@{\hspace{1ex}}r@{\hspace{1ex}}r@{\hspace{1ex}}r@{\hspace{1ex}}} 
36&20&15&20&35\\
20&36&30&20&20\\
15&30&36&30&15\\
20&20&30&36&20\\
35&20&15&20&36
\end {array}\right]
\end{array}
\end{displaymath}\\[2ex]

The transformation matrix $M$ from spherical to real harmonics is
\begin{displaymath}
\begin{array}{c@{\hspace{5ex}}c@{\hspace{6ex}}cc@{\hspace{5ex}}c@{\hspace{6ex}}c}
\quad & \displaystyle M=\frac{1}{\sqrt2}
\left[
\begin {array}{r@{\hspace{1ex}}r@{\hspace{1ex}}r@{\hspace{1ex}}r@{\hspace{1ex}}r@{\hspace{1ex}}} 
- i&0&0&0&i\\ 
       0&i&0&i&0\\ 
       0&0&\sqrt2&0&0\\ 
       0&-1&0&1&0\\ 
       1&0&0&0&1\\ 
\end {array}
\right]. 
\end{array}
\end{displaymath}\\[1ex]

The Coulomb integrals can also be expressed as linear combination of Racah parameters; for the $d$ shell the latter are
\begin{eqnarray*}
A&=&F_0-\frac{49}{441}F_4 \quad \quad
B=\frac{1}{49}F_2-\frac{5}{441}F_4   \quad \quad
C=\frac{35}{441}F_4.
\end{eqnarray*}


\end{document}